\documentclass[aps,prl,reprint,superscriptaddress,showpacs]{revtex4-1}

\usepackage{graphicx}
\usepackage{color}
\usepackage{booktabs}
\usepackage{bm}
\usepackage{longtable}
\usepackage{amsmath}
\usepackage{dcolumn}
\usepackage{placeins}
\usepackage{expdlist}

\begin{document}

\title {Parametric amplification and noise-squeezing in room temperature atomic vapours}

\author{V. Guarrera} 
\email{v.guarrera@bham.ac.uk}
\affiliation{Midlands Ultracold Atom Research Centre, School of Physics and Astronomy, University of Birmingham,
Edgbaston, Birmingham B15 2TT, United Kingdom}
\affiliation{National Physical Laboratory, Hampton Road, Teddington, TW11 0LW, United Kingdom}
\author{R. Gartman}
\affiliation{National Physical Laboratory, Hampton Road, Teddington, TW11 0LW, United Kingdom}
\author{G. Bevilacqua} 
\affiliation{DIISM, Universit\`a di Siena, via Roma 56, 53100 Siena, Italy}
\author{G. Barontini} 
\affiliation{Midlands Ultracold Atom Research Centre, School of Physics and Astronomy, University of Birmingham,
Edgbaston, Birmingham B15 2TT, United Kingdom}
\author{W. Chalupczak}
\affiliation{National Physical Laboratory, Hampton Road, Teddington, TW11 0LW, United Kingdom}

\date{\today}

\begin{abstract}
We report on the use of parametric excitation to coherently manipulate the collective spin state of an atomic vapour at room temperature. Signatures of the parametric excitation are detected in the ground-state spin evolution. These include the excitation spectrum of the atomic coherences, which contains resonances at frequencies characteristic of the parametric process. The amplitudes of the signal quadratures show amplification and attenuation, and their noise distribution is characterized by a strong asymmetry, similarly to those observed in mechanical oscillators. The parametric excitation is produced by periodic modulation of the pumping beam, exploiting a Bell-Bloom-like technique widely used in atomic magnetometry. Notably, we find that the noise-squeezing obtained by this technique enhances the signal-to-noise ratio of the measurements up to a factor of 10, and improves the performance of a Bell-Bloom magnetometer by a factor of 3. 
\end{abstract}

\maketitle

\textit{Introduction.} 
Parametric excitation provides an excellent tool for gaining control over signal amplitude and noise in oscillating systems. As it is a coherent process with phase-sensitive character, it can be used to generate squeezed states that are crucial for technological developments, including implementations of quantum technology. Proof-of-principle demonstrations of this effect have been achieved in a variety of different systems ranging from mechanical oscillators \cite{rugar1991,pritchard1992,Harris2013,Szorkovszky2011,Pontin2014}, to optical modes \cite{slusher1995} and ultracold atoms \cite{Hoang2016}. Parametric excitation of an oscillator is obtained by periodically modulating one of the system parameters, such as the oscillator natural frequency $\omega_0$ or the damping rate, at frequencies which are sub-multiples of $2\omega_0$ \cite{Landau}. The excitation occurs because of a distortion of the system orbits in the phase-space, without involving any direct perturbation of the equilibrium position of the oscillator. As a result, the dynamic response of the system at $\omega_0$ is enhanced. It also shows a characteristic phase dependence: while the quadrature of the motion in phase with the parametric modulation is amplified, the orthogonal (out-of-phase) quadrature is attenuated. Likewise, fluctuations of the two quadratures are modified. This technique can be exploited to considerably reduce the uncertainty in the determination of the frequency of an oscillator \cite{pritchard1992}, and to achieve noise reduction below both classical (e.g. thermal or technical) \cite{rugar1991} and standard quantum limits \cite{Hoang2016,wollman2015}. 

\begin{figure}[h!]
\includegraphics[width=\columnwidth]{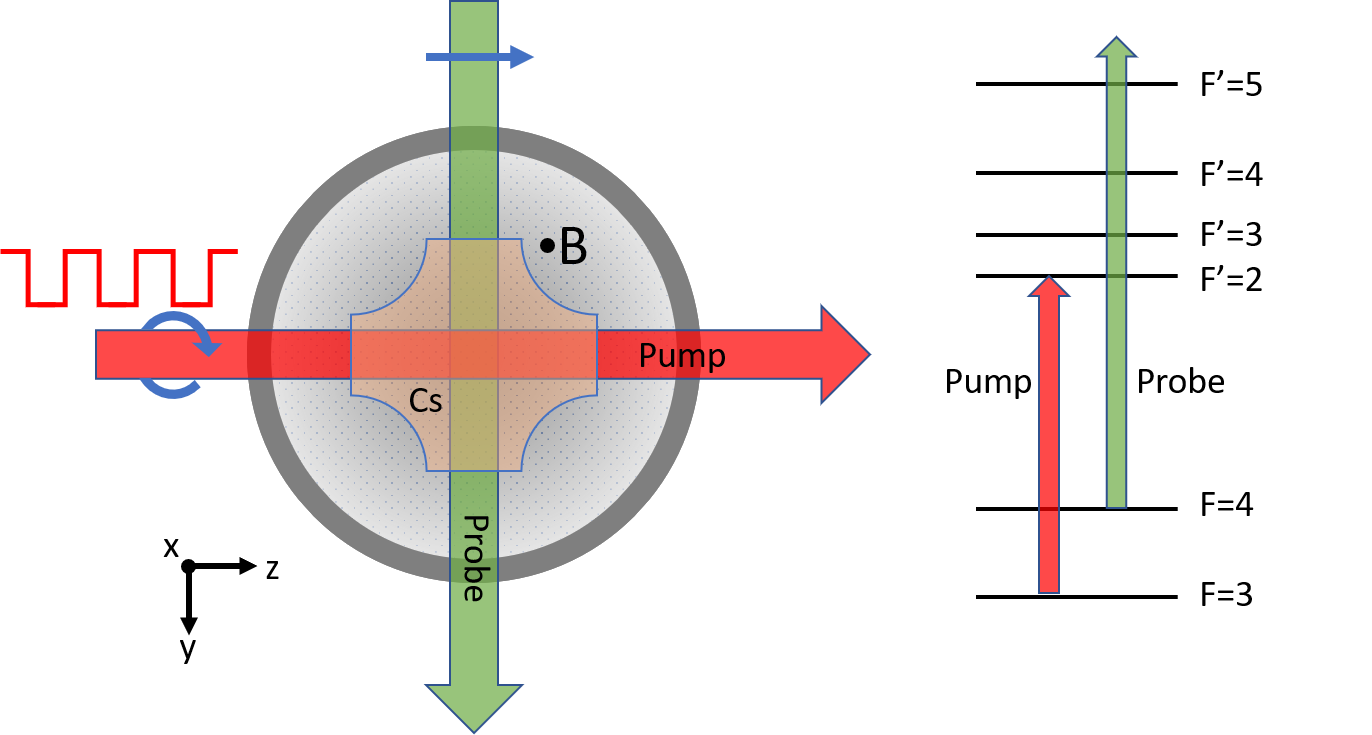}
\caption{Schematics of the apparatus: a circularly polarized amplitude-modulated pump beam creates atomic coherences in the ground-states of the Cs atoms by resonant (F=3) and spin-exchange pumping (F=4). The generated collective atomic spin precesses orthogonally to the applied static magnetic field B. A linearly-polarized probe beam is used to extract information on the atomic precession via Faraday-like, non-destructive measurements. }
\label{fig:Setup}
\end{figure}

Here we demonstrate parametric amplification and noise-squeezing in the collective atomic spin system realized by an ensemble of optically-pumped room temperature atoms, which is a widely exploited platform for magnetometry \cite{allred2002}, quantum information processing \cite{julsgaard2004}, and rotation sensing \cite{limes2018}. We show that by using such parametric squeezing we can enhance the signal-to-noise ratio in a Bell-Bloom magnetometer by up to a factor of 10. To understand the working principle of our parametric effect, let us consider a collective atomic spin precessing around a dc magnetic field $\textbf{B}=B \hat{x}$ (see Fig.~\ref{fig:Setup}) with natural (Larmor) frequency $\omega_L= g B $, where $g$ is the gyromagnetic ratio of the atoms. This system can be mapped onto an harmonic oscillator with frequency $\omega_0=\omega_L$, where the components of the oscillating spin $\widehat{F}_y$ and $\widehat{F}_z$ take the place of position $\widehat{X}$ and momentum $\widehat{P}$ of a mechanical oscillator \cite{Polzik2015}. Parametric amplification and squeezing might be achieved in atomic spin systems by periodic modulation of either $\omega_L$, via modulation of the magnetic field amplitude, or the damping rate $\gamma$ of the Larmor precession. In this Letter, we investigate this latter case by modulating the damping rate via the intensity modulation of the pump beam in a Bell-Bloom magnetometer \cite{bellbloom}. 

In contrast to the standard Bell-Bloom scheme, in the present work we use a relatively strong pump resonantly coupled to one of the atomic ground-states of caesium atoms, as shown in Fig.~\ref{fig:Setup}. The pump term, in this case, can be written in the density matrix formalism as \cite{Gartman2018}:
\begin{align}
\Gamma(t)  &=- L_R \left\lbrace WW^{\dagger}, \rho \right\rbrace  +2 L_R W \rho_{e} W^{\dagger} 
+ (\rho_e')_{sp}  \\
& =\Gamma_a+\Gamma_p \rho \nonumber
\label{eq:pompa}
\end{align}
where $W=-E_0(t) \Pi \textbf{d} \cdot \textbf{e}^* \Pi_e$ ($W^{\dagger}=E_0(t) \Pi_e \textbf{d} \cdot \textbf{e} \Pi$), with $E_0(t)$ and $\textbf{e}$ the amplitude and polarization versor of the laser field respectively, and $\textbf{d}$ the induced atomic dipole moment. $\Pi$ and $\Pi_{e}$ are the projectors on the ground and excited state (labelled with the letter $e$) which are connected by the pump, and $L_R$ depends on the natural linewidth of the transition. The first term in Eq.~1 accounts for depopulation of the ground-state, while the second and third refer to repopulation due to stimulated and spontaneous emission. In the dynamics of the macroscopic spin, the power broadening due to the pump beam contributes with a parametric term $\Gamma_p$, while the depopulation and repopulation due to stimulated and spontaneous emission are taken into account by an additive term $\Gamma_a$. The light shift induced by the pump is neglected in the condition of resonant pumping. More explicitly, we can write the evolution of the spin components of interest as: 
\begin{align}
& \frac{d^2 F_z}{dt^2}+ (\gamma+\Gamma_p(t)) \frac{d F_z}{dt}+\omega_L^2 F_z -\omega_L\Gamma_a(t) =0  \\
& \frac{d F_y}{dt}=-\omega_L F_z \nonumber
\label{eq:eq of motion}
\end{align}   
where $\gamma \sim \frac{1}{T_2}$ is the rate of spin coherence relaxation due to spin-exchange collisions, and the pump is applied along the $\widehat{z}$-axis. Note that Eq.~2 describes the dynamics for the atoms in the directly pumped ground-state. However, polarization and coherence generated by the pumping process in this state are also transferred to the other ground-state by spin-exchange interactions (indirect pumping), which consists of a term proportional to $\vert \textbf{F} \vert$ \cite{Gartman2018}. Hence Eq.~2 is also valid for the atomic dynamics in the indirectly pumped ground-state. This simplification of the model allows us to easily extract useful information on the dynamical evolution of the atomic system.   

From the equations above, it follows that in the presence of a modulated additive pump ($\Gamma_p \sim 0$) coherent oscillation at the Larmor frequency has a transient character, while coherent oscillation at the modulation frequency $\omega_M$ takes over in the stationary regime \cite{Grujic2013}. In other words, the pump acts on the collective spin system like a forcing term on a damped harmonic oscillator. For a relatively strong pump ($\Gamma_p \gg0$), while these features remain generally valid, we have already observed that the coherence spectrum becomes more complex \cite{Gartman2018}.     

\textit{Experimental setup.} Details of the experimental configuration are described in \cite{Gartman2015} and here we only recall the key elements. As shown in Fig.~\ref{fig:Setup}, a caesium vapour is held in an anti-relaxation, paraffin coated, cross-shaped glass cell, placed within five layers of mu metal shields which eliminate ambient magnetic fields. All measurements are performed with an atomic density of $0.33 \times10^{11} \text{cm}^{-3}$, at room temperature. The pumping is done by using a circularly polarised laser beam, frequency locked to the caesium $6\,^2$S$_{1/2}$ F=3$\rightarrow{}6\,^2$P$_{3/2}$ F'=2 transition. The pump intensity is periodically switched on and off following a square modulation function. The signal produced by the $F=4$ ground-state atomic coherences is read out by a probe beam propagating along a direction orthogonal to the pump beam, frequency locked to the $6\,^2$S$_{1/2}$ F = 4$\rightarrow{}6\,^2$P$_{3/2}$ F' = 5 transition, and subsequently frequency shifted by $960$ MHz to the blue side. Probing the $F=4$ ground-state allows to safely decouple the effects of the pump and probe beams, while still capturing features of the atomic polarization dynamics of the directly pumped state. 

A pair of Helmholtz coils produces a magnetic field orthogonal to the pump and probe beams. In this configuration, the continuous Faraday-type polarization rotation measurements provide information on the component of the collective spin parallel to the probe beam $S(t)=k \langle F_y(t) \rangle$, where $k$ depends on the details of the atomic transition used for probing and on the geometry of the magnetometer \cite{takahashi1999}. The probe light transmitted through the cell is analysed by a polarimeter and processed by a signal analyser or by a lock-in amplifier. This latter allows the extraction of the spin projections $(s_c,s_s)$ in a frame rotating at frequency $\omega_L$, so that $S(t)=s_c(t) \cos(\omega_L t) + s_s(t)  \sin(\omega_L t)$, which have slow fluctuations. 
\begin{figure}
\centering
\includegraphics[width=\columnwidth]{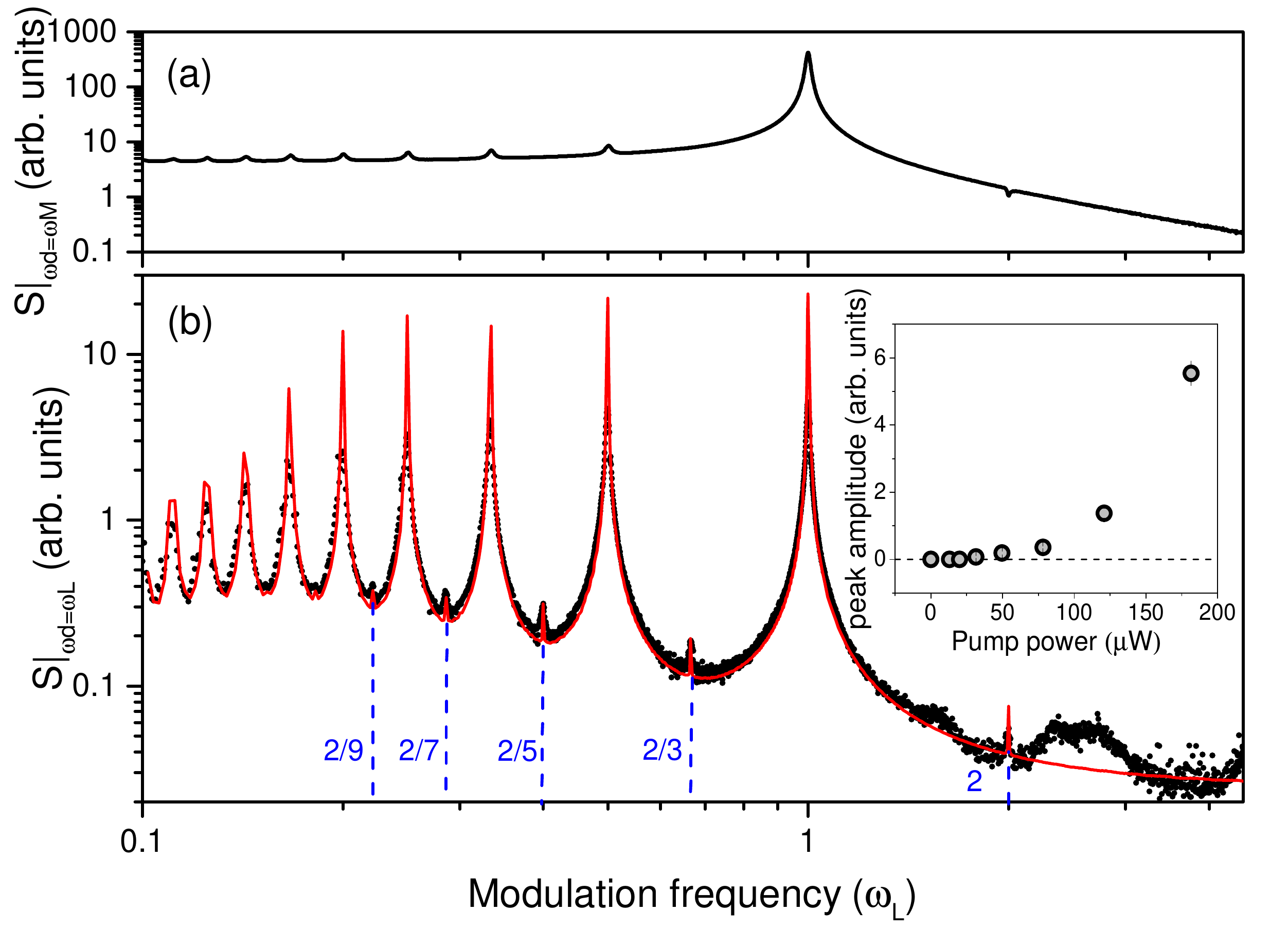}
\caption{ (color online) Amplitude dependence of the $F=4$ atomic coherences oscillating at (a) $\omega_M$, and (b) $\omega_{\text{L}}$ on the pump beam amplitude modulation frequency $\omega_M$, recorded with a pump beam power of $78 \mu$W. All measurements were made with a probe beam power of 100 $\mu$W, and $\omega_L/2\pi=270$ Hz. The solid red line represents the signal obtained by numerically solving Eq.~2. The modulation frequency is expressed in units of $\omega_L$. In the inset the amplitude of a parametric peak ($ \omega_M= 2/3 \omega_L $) is shown as a function of the pump beam power.} 
\label{fig:Spectrum}
\end{figure}
  
\textit{Excitation spectrum.} 
Figure~\ref{fig:Spectrum}(a) shows the excitation spectrum of the component at $\omega_M$ in our configuration, which reveals the typical response of a harmonic oscillator driven by pulsed excitation. Since the amplitude of the pump beam is modulated with a square pulsed waveform, the spectrum of the optical excitation consists of a series of harmonics \cite{Grujic2013, Gartman2015}. A small feature is also visible at $2 \omega_L$ which is associated with alignment produced by the pump beam for higher pump powers \cite{budkerbook}. Up to this point, our system shows no difference with respect to a standard Bell-Bloom scheme. 

To take a closer look at the dynamical response of the system, we extract the amplitude of the coherence signal oscillating at $\omega_{\text{L}}$ from the FFT of the whole magneto-optical rotation signal. This amplitude is shown for different modulation frequencies in Fig.~\ref{fig:Spectrum}(b). We observe that, besides the usual Bell-Bloom resonances, a series of resonant features are visible at the characteristic frequencies of parametric amplification, i.e. $n  \omega_M = 2 \omega_{\text{L}}$. On the basis of the above observations, these correspond to an enhancement of the detected signal oscillating at $\omega_L$ that is not induced by pumping synchronization with the spin free evolution. The parametric peaks are visible for pump powers above $20 \mu$W, and their amplitude progressively increases for higher values of the pump, as shown in the inset of Fig.~\ref{fig:Spectrum}(b). We have also verified that they do not depend on the relative orientation of the probe beam and magnetic field for subtended angles between $45^ \circ $ and  $90^ \circ $ \cite{footnote}.  

We find good agreement between the experimental data and numerical simulations performed by solving Eq.~2, where $\Gamma_p/\Gamma$ is kept as a free parameter, and $\gamma$, $\omega_L$, and the time-averaged $\Gamma_p$ are measured experimentally, see Fig. \ref{fig:Spectrum}(b). We observe that the amplitude of the signal at modulation frequencies $2/(2n+1) \omega_L$ is smaller than that at sub-multiples of $\omega_L$. This is due to the parametric amplification being effective only during the initial transient of the oscillation at $\omega_L$, when the system is not in an established auto-oscillating regime.  

\begin{figure}[h!]
\includegraphics[width=\columnwidth]{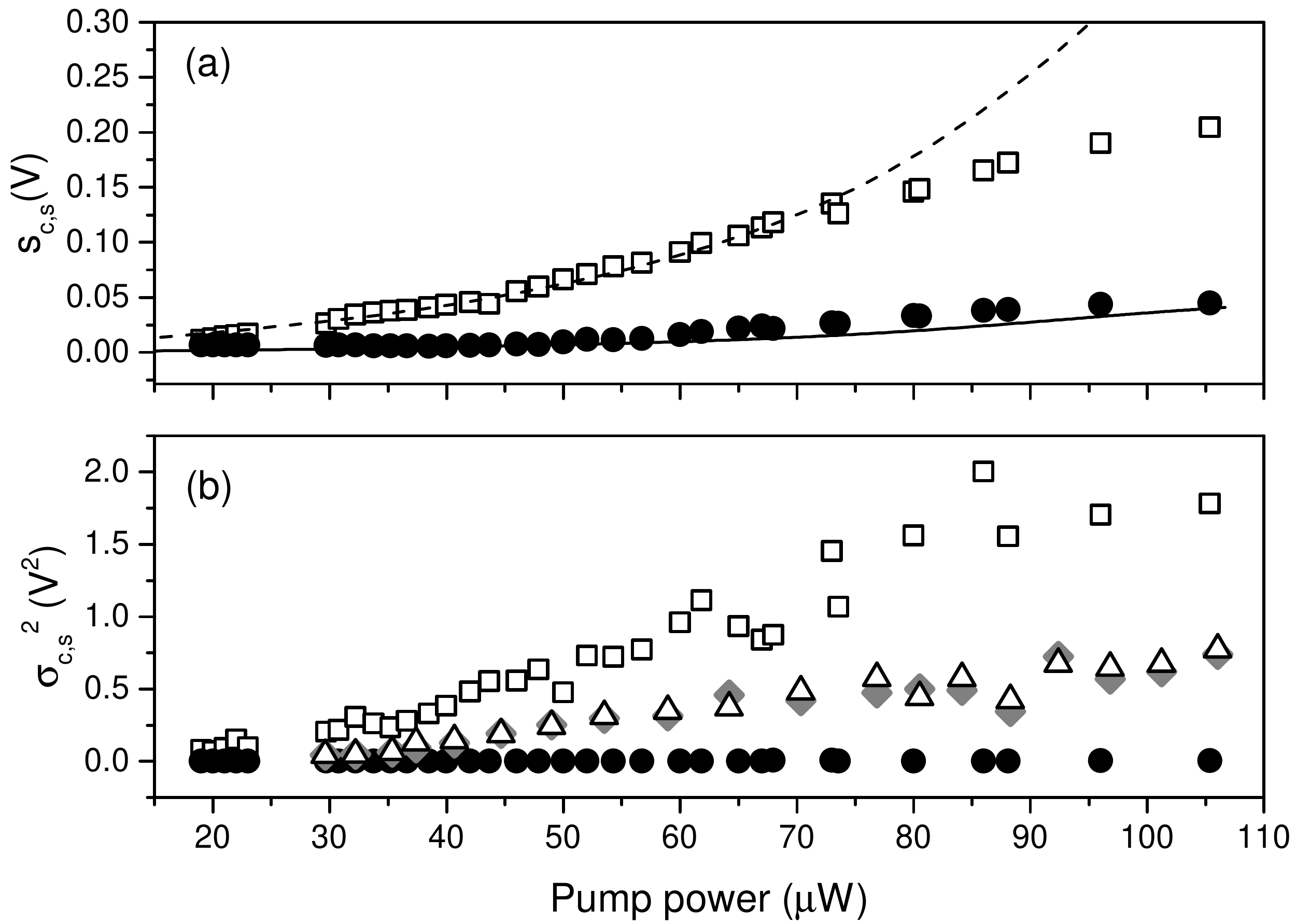}
\caption{(a) Amplitude of the in-phase (open squares) and out-of-phase (filled dots) components of the magneto-optical rotation signal oscillating at frequency $\omega_L$ as a function of the pump beam power. The modulation frequency for these measurements is set to $2\omega_L$. The probe beam power is 120 $\mu$W. Solid lines are numerically calculated for the same parameters of Fig.~\ref{fig:Spectrum}(b). (b) Variance of the quadrature amplitudes over $350$ measurements for the in-phase (open squares) and out-of-phase (filled dots) components as a function of the pump power. For comparison, the variance measured with a modulation frequency of $4 \omega_L$ shows no significant asymmetry between the two quadratures' components (open triangles and gray diamonds).}
\label{fig:Amplitude}
\end{figure}

\noindent \textit{Parametric amplification and squeezing at $\omega_M=2 \omega_L$.} First, we concentrate our analysis on the behaviour of the peak that appears in the spectrum for $\omega_{\text{M}} = 2 \omega_{\text{L}}$. On the one hand, this allows us to address a characteristic parametric feature of the spectrum, on the other it allows us to single out the purely parametric effect as the natural frequency of the oscillation and the modulation frequency do not overlap. The amplitudes of the quadrature components $(s_c,s_s)$ in and out-of-phase with respect to the external modulation are extracted from the excitation spectrum measured 130 ms after the beginning of the pumping process, see Fig.~\ref{fig:Amplitude}(a). The lock-in amplifier used in these measurements is referenced to $\frac{1}{2} \omega_M$, with $\omega_M/2\pi=2 \times 270$ Hz, while the Larmor frequency is scanned across the resonance $\omega_L=\frac{1}{2} \omega_M$. Figure~\ref{fig:Amplitude}(a) shows the dependence of the amplitude of the two signal quadratures on the pump power. Unlike non-parametric excitation, for which increasing the pump power below saturation leads to enhancement of overall polarization of the atomic system with the two quadratures growing by the same amount, here we observe that the in-phase and out-of-phase components grow at very different rates. In particular, the out-of-phase (in-phase) signal component is attenuated (amplified) by the parametric excitation with respect to the overall polarization growth. This can be understood by analytically solving the dynamics for the quadrature components for a simple sinusoidal modulation of the pump. One obtains $s_s(t)=s_0 e^{-\frac{\gamma}{2}t} e^{-\frac{\Gamma_p}{4}t}$ and $s_c(t)=s_0 e^{-\frac{\gamma}{2}t} e^{\frac{\Gamma_p}{4}t}$, where $s_0$ is the maximum atomic polarization amplitude, also a function of the pump power. Comparison with numerical solutions of Eq.~2 provides good agreement for pump powers up to roughly $70 \mu$W \cite{footnote2}. As discussed above, phase sensitive gain is a characteristic feature of a parametric amplification process \cite{rugar1991}. 

\begin{figure}[h!]
\includegraphics[width=\columnwidth]{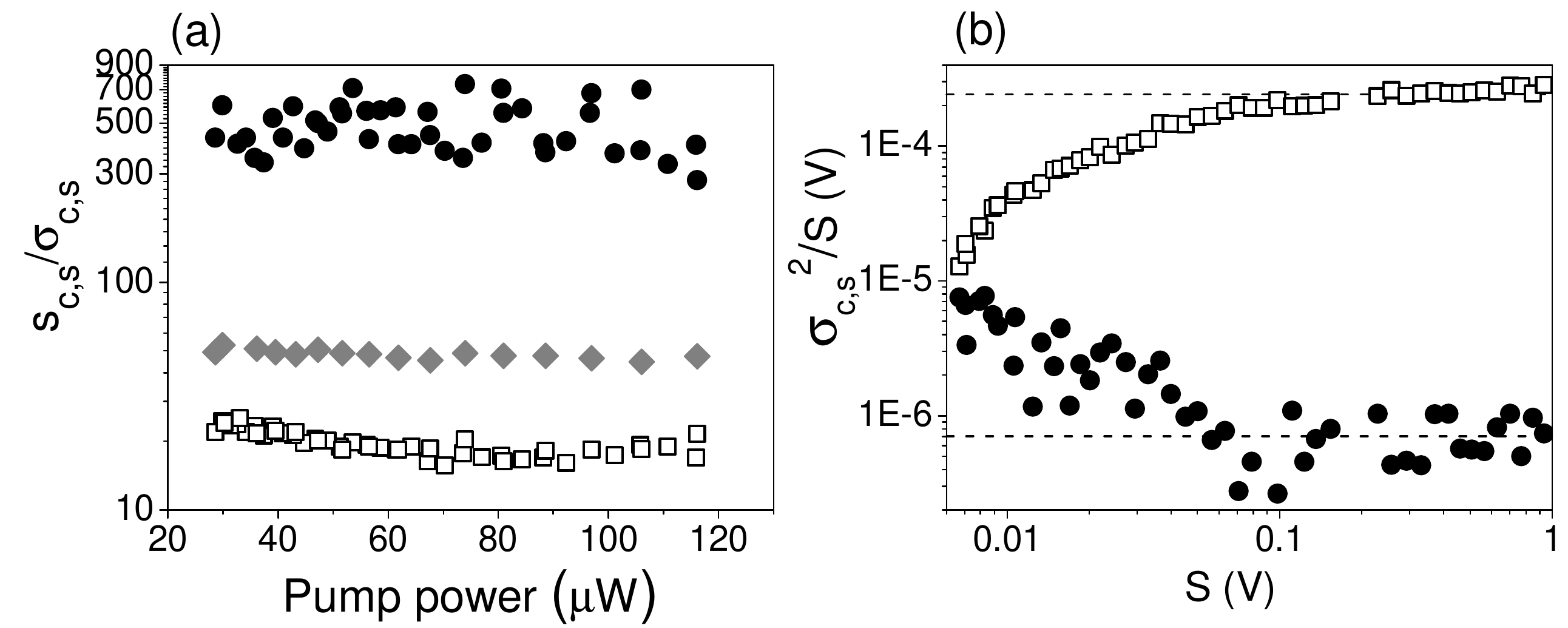}
\caption {(a) Signal-to-noise ratio (SNR) of the signal quadratures measured for $\omega_M=2\omega_L$ as a function of the pump power. The out-of-phase quadrature (filled dots) shows an improved SNR with respect to the in-phase quadrature (open squares), and to the case of no modulation (grey diamonds). (b) Noise scaling with signal amplitude. As far as the parametric effect plays a role, the in (out-of) phase quadrature scales faster (slower) than linear with respect to the signal amplitude. For larger pump powers, this effect ceases and the noise scales linearly, compatibly with the atomic-shot noise limit.}
\label{fig:Squeezing}
\end{figure}
The effect of the parametric amplification on the two quadratures is also reflected in the behaviour of the relative noise components. We observe that with increasing pump powers, the noise is strongly squeezed along the attenuated quadrature and expanded along the other, see Fig.~\ref{fig:Amplitude}(b), leading to an asymmetric noise distribution. For comparison, in Fig.~\ref{fig:Amplitude}(b) we also show the measurement of the quadratures' noise for the non-parametric $\omega_M=4 \omega_L$. One can see that in the latter case the noise is equally distributed between the two quadratures. 

Amplitude fluctuations are important as they affect the determination of the natural frequency of an oscillator in the presence of noise, and reduce the precision of magnetic field measurements. To evaluate this parameter we plot in Fig.~\ref{fig:Squeezing}(a) the signal-to-noise ratio (SNR) obtained by processing the polarization rotation signal with the lock-in amplifier as a function of the pump power for modulation frequency $2\omega_L$. The out-of-phase component (filled dots) shows an increased SNR with respect to the in-phase component (open squares) and, most importantly, to any component in the absence of modulation of $\gamma$ (gray diamonds). Notably, with respect to this latter case the SNR enhancement is of 10 dB. We further analyse the noise scaling with the signal amplitude as shown in Fig.~\ref{fig:Squeezing}(b). At lower signal amplitudes, corresponding to pump powers smaller than $60 \mu$W, the noise of the amplified (attenuated) component scales faster (slower) than linear. A fit with a function $\sigma=as^b$ to the measured data yields $b=1.52(3)$ and $b=0.44(6)$ respectively. A linear scaling, consistent with projection-noise-limited measurements appears for higher signal amplitudes, consistently with the deviation from the parametric behaviour displayed in Fig.~3(a). 

\noindent \textit{Enhanced Bell-Bloom magnetometer.}
Having analysed the main features of the parametric excitation, we now show how such excitation improves the performance of a Bell-Bloom magnetometer beyond atomic projection noise. To this end, we report in Fig.~5 the SNR measured at $\omega_M=\omega_L$ as a function of the pump power. We observe that with a pump power of $45$ $\mu$W the SNR of our magnetometer is improved by a factor of 3. We attribute the reduced enhancement with respect to the case at $2\omega_L$ to the presence of additional noise introduced by the pump which cannot be filtered out in the FFT of the signal. Nonetheless, our measurements reveal that the parametric noise-squeezing obtained in the attenuated quadrature is a very robust technique for reducing the noise and improving SNR independently of our system being limited by atomic projection-noise (for the case $\omega_M=2 \omega_L$) or technical noise (for the case $\omega_M= \omega_L$), as the method is fully classical and does not rely on the quantum features of the system. Currently, our system's performance for $\omega_M= \omega_L$ at low-frequencies is affected by Johnson noise of the magnetic shields. The presented technique allows an improvement in atomic noise to roughly 15 fT/$\sqrt(Hz)$, and, with an improved shielding, has the potential to bring the sensitivity below the 1 fT/$\sqrt(Hz)$ level \cite{witold2012}. We finally note that the continuous measurement performed on the atomic system has minimal impact on the noise distribution. At the parametric resonances, we find that the degree of asymmetry in the noise distribution weakly depends on the probe beam power, and increases by roughly $10 \%$ when the power is raised from $20$ to $500$ $\mu$W while the noise axis orientation does not significantly change in this range.  
\begin{figure}[t!]
\includegraphics[width=0.8\columnwidth]{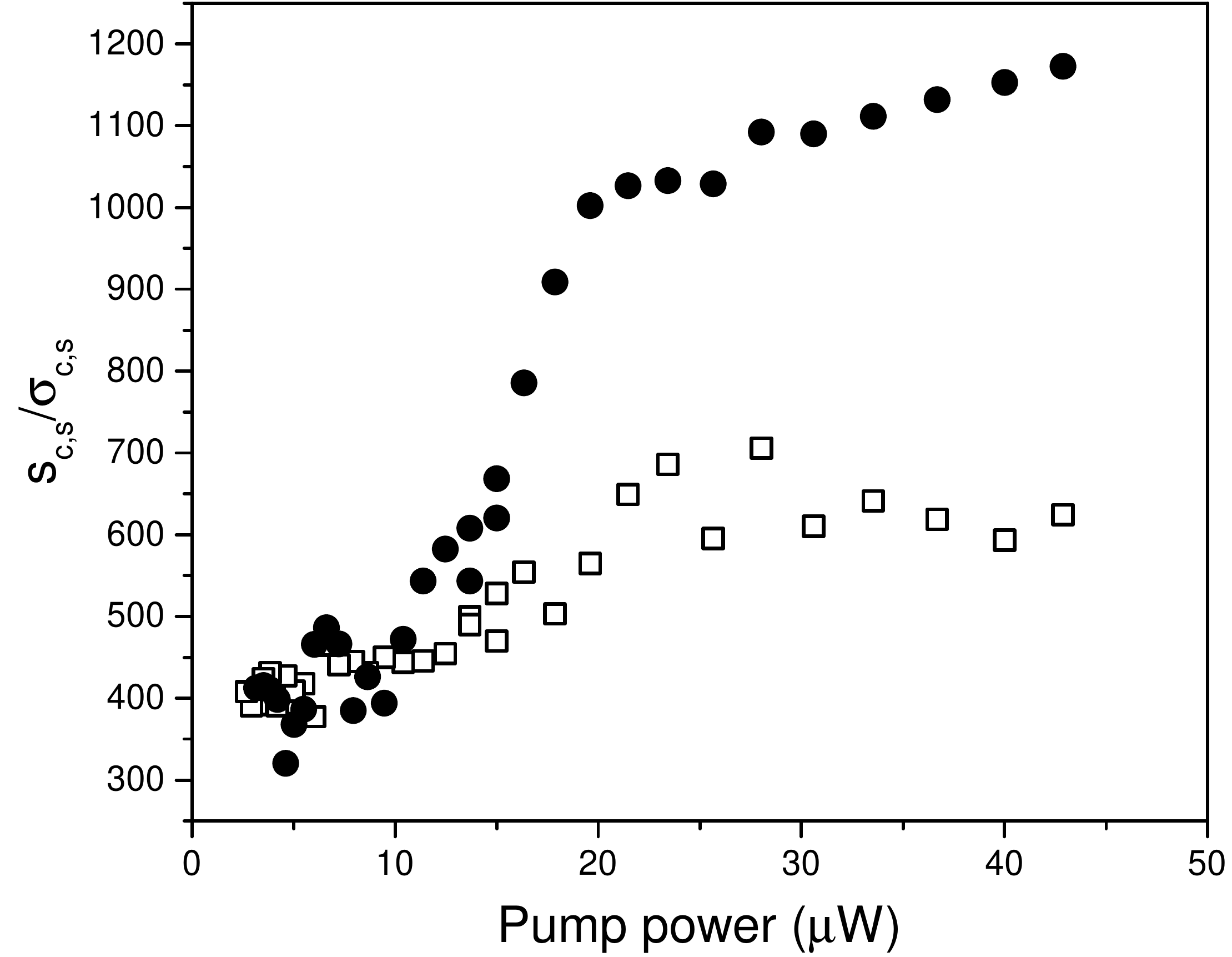}
\caption {Signal-to-noise ratio of a Bell-Bloom magnetometer, as a function of the pump power. Similarly to the case of modulation at $2\omega_L$, the out-of-phase quadrature (filled dots) is increased with respect to the in-phase quadrature (open squares) for powers larger than roughly $20$ $\mu$W. Note that, as expected the overall magnetization is higher when hitting the Bell-Bloom resonance with respect to other modulation frequencies.}
\label{fig:Squeezing2}
\end{figure}

\textit{Conclusions.} In conclusion, we have investigated the transient dynamics induced by a periodic modulation of the pump beam power in a Bell-Bloom magnetometer. Signatures of parametric amplification have been detected in the coherences of the optically pumped Cs vapour. First parametric resonances are visible in the excitation spectrum, which would not appear for a simply periodically driven oscillator. Parametric amplification and attenuation is observed in the in-phase and out-of-phase quadrature components respectively, which reflects in strong asymmetry in their amplitude noise distribution. Parametric amplification and squeezing provide a robust technique to manipulate the signal and noise distribution in a magnetometer for precision measurements. We have indeed shown that the signal-to-noise ratio can be increased up to a factor of 10 when the modulation frequency hits a parametric resonance and that the performance of a Bell-Bloom magnetometer can be enhanced by a factor of 3. Moreover, the system can be eventually led into a self-oscillating regime, which might be used to increase the measurement interrogation time beyond its natural decoherence and measurement bandwidth. 
Finally we foresee application for noise control in combination with non-demolition continuous measurements, as already explored in mechanical systems, for "bright" squeezing generation \cite{Pontin2014}.

\begin{acknowledgements}
The work was funded by the UK Department for Business, Innovation and Skills as part of the National Measurement System Programme. V.G. was supported by the EPSRC (Grant No. EP/S000992/1). We would like to thank R. Hendricks for critical reading of the manuscript. 
\end{acknowledgements}


\begin{thebibliography}{10}
\bibitem{footnote} We detect also a few broad features for modulation frequencies higher than $2 \omega_L$ which have no direct impact on the present study and will be addressed in a separate paper.
\bibitem{footnote2} Our simple model does not take into account contribution from Stark-shift and saturation effects from the pump which become more important for higher pump powers.
\end{thebibliography}
\end{document}